\documentclass[11pt]{article}
\usepackage{ACL2023}
\usepackage{algorithm}
\usepackage{algpseudocode}

\usepackage{times}
\usepackage{latexsym}
\usepackage[T1]{fontenc}
\usepackage[utf8]{inputenc}
\usepackage{microtype}

\usepackage{lipsum}
\usepackage{url}
\usepackage{graphicx}  %Required
\usepackage{CJK}
\usepackage{bm}
\usepackage{float}
\usepackage{xcolor,colortbl}
\usepackage{adjustbox}
\usepackage{subfigure}
\usepackage{booktabs,multirow}
\usepackage{bigstrut}
\usepackage{paralist}
\usepackage{bbm}
\usepackage{makecell}
\usepackage{nicefrac}       % compact symbols for 1/2, etc.
\usepackage{microtype} 
\usepackage{epsfig}
\usepackage{diagbox}
\usepackage{framed}
\usepackage{empheq}
\usepackage{array}
\usepackage{enumitem}
\usepackage{mdwlist}
\usepackage{xspace,mfirstuc,tabulary}
\usepackage{tcolorbox}
\usepackage{placeins}
\usepackage{microtype}

\usepackage[normalem]{ulem}
\useunder{\uline}{\ul}{}
\usepackage{threeparttable}
\usepackage{makecell}
\usepackage{booktabs}

\usepackage{xspace}
\usepackage{url}

\usepackage{graphicx}
\usepackage{subfigure}
\usepackage{multirow}
\usepackage{multicol}
\usepackage{pifont}
\usepackage{enumitem}
\usepackage{appendix}
\usepackage[table]{xcolor}
\usepackage{colortbl}
\usepackage{graphicx}
\usepackage{multirow}
\usepackage[table,xcdraw]{xcolor}

\usepackage{tikz}
\newcommand*{\circled}[1]{\lower.7ex\hbox{\tikz\draw (0pt, 0pt)%
    circle (.5em) node {\makebox[1em][c]{\small #1}};}}
\usepackage[utf8]{inputenc}
\usepackage[lined,vlined,ruled,linesnumbered,algo2e]{algorithm2e}
\newcommand{\hlt}[1]{\textcolor{blue}{#1}}
\newcommand{\para}[1]{\paragraph{\textnormal{\textbf{#1}:}}}
\newcommand{\uls}{\begin{itemize}[leftmargin=*]}
\newcommand{\ule}{\end{itemize}}
\newcommand{\ols}{\begin{enumerate}[leftmargin=*]}
\newcommand{\ole}{\end{enumerate}}
\newcommand{\li}{\item}

\begin{document}

\author{
	Manojit Chakraborty\textsuperscript{1} \quad 
	Madhusudan Ghosh\textsuperscript{2} \quad
	Rishabh Gupta\textsuperscript{1} \\
	\textsuperscript{1}Bosch Research \& Technology Center, India \\
	\textsuperscript{2}Indian Association for the Cultivation of Science, Kolkata \\
	\texttt{Manojit.Chakraborty@in.bosch.com} \quad
	\texttt{madhusuda.iacs@gmail.com} \quad \\
	\texttt{Gupta.Rishabh@in.bosch.com}
}

%%g


% \title{Cost-Efficient Long Code Translation: Leveraging Identifier Replacement for Code Translation using Large Language Models}
\title{LLM Based Long Code Translation using Identifier Replacement}
\maketitle

\begin{abstract}
% In recent years, Large Language Models (LLMs) have demonstrated impressive capabilities in handling a wide range of natural language processing tasks without task-specific training, making them effective zero-shot predictors.
In the domain of software development, LLMs have been utilized to automate tasks such as code translation, where source code from one programming language is translated to another while preserving its functionality. However, 
LLMs often struggle with long source codes that don't fit into the context window, which produces inaccurate translations. To address this, we propose a novel zero-shot code translation method that incorporates identifier replacement. By substituting user-given long identifiers with generalized placeholders during translation, our method allows the LLM to focus on the logical structure of the code, by reducing token count and memory usage, which improves the efficiency and cost-effectiveness of long code translation.
Our empirical results demonstrate that our approach preserves syntactical and hierarchical information and produces translation results with reduced tokens.
% in compared to existing approaches.
% \keywords{LLMs, Long Code Translation, Cost Savings, Token Length}
% After translation, the placeholders are mapped back to their original identifiers to ensure the functional correctness of the translated code. 
\end{abstract}




\section{Introduction}
\label{sec:intro}
Large Language Models (LLMs) demonstrate impressive abilities in capturing text semantics abstractly, without the need for task-specific training \cite{radford,gpt3,arora2023ask,weidinger2022taxonomy}. This characteristic of ``general intelligence" enables LLMs to function as zero-shot predictors across a variety of downstream tasks, including question answering \cite{li-etal-2023-shot}, document retrieval \cite{pradeep2023does}, and code translation \cite{Pan_2024}. 
However, the practical application of LLMs in software engineering faces several significant challenges. One notable area is code translation, which involves converting large, complex codebases from one programming language to another. This process is essential for organizations aiming to migrate legacy systems or to integrate various programming languages within large code repositories\cite{ahmad-etal-2023-avatar, pan2024lost, krishna2021transforming, nitin2022cargo}. It is also critical to ensure that the functional and logical integrity of the code is maintained throughout the translation process \cite{ahmad-etal-2023-avatar, pan2024lost}. 
But the inherent context length limit of every LLM poses a significant challenge for code translation tasks \cite{radford,gpt3}. Long code bases frequently contain intricate dependencies, and this limitation often necessitates partitioning the code into smaller segments. This division complicates the translation process and increases the risk of losing structural consistency across different sections of the code \cite{jana2023attention,gong2023adelt}.

Source code identifiers, particularly complex user-defined function names, class names, and variable names, play a crucial role in structuring large codebases. These identifiers often contain descriptive metadata or project-specific conventions, leading to increase token usage when processed by LLMs. While identifiers improve code readability for human developers, it can contribute to rapid token consumption in LLMs models, reducing the available context window for performing code translation. 
% Additionally, identifiers are key to preserving dependencies across different functions and modules. During translation, inconsistencies in identifier mappings can disrupt the logical flow of the translated code, leading to errors in function calls or object references.
To alleviate the above challenges, we present a novel zero-shot code translation method that employs an \textit{identifier replacement}\footnote{https://anonymous.4open.science/r/LCT-LLM/} strategy to optimize the translation process using LLMs. While token reduction is the most immediate benefit of our identifier replacement strategy, a subtler but equally important reason for improved accuracy is the shift in the model's attention distribution. Long and descriptive identifiers often dilute the attention of LLMs across semantically redundant sub-tokens. By compressing these identifiers into compact placeholders (e.g., \texttt{id\_1}), the model is nudged to prioritize the underlying syntactic and control-flow structure of the program rather than overfitting to identifier surface forms. This mechanism effectively reduces noise and enables the LLM to better capture execution-relevant relationships (e.g., function calls, data dependencies) during translation. In other words, identifier replacement transforms the problem into a more syntax-driven translation task, which explains the observed accuracy improvements for procedural languages in our experiments. This tailored extraction process refines the preprocessing stage by differentiating between critical identifiers, such as function and class names, and non-essential ones, thereby enhancing translation fidelity. By simplifying the task for the LLM and prioritizing the logical structure of the code over lengthy identifiers, our approach enables the model to manage longer code sequences within its token limit more effectively~\cite{jana2023attention} ~\cite{dinh2023largelanguagemodelscode}. This not only decreases the number of tokens needed to represent the code but also significantly reduces the computational cost of inference.

% Unlike conventional preprocessing strategies which primarily focus on lexical tokenization \cite{tree_sitter}, our approach employs a \textit{context-aware identifier extraction strategy}, leveraging a custom grammar-based filtering mechanism that ensures preservation of essential code semantics.

% This technique involves dynamically substituting lengthy identifiers with compact placeholders prior to translation and restoring them afterward. 

% More specifically speaking, in this method, complex, long user-given identifiers are replaced with short generalized placeholders during the translation process. 

% By focusing on the logical structure of the code rather than these long identifiers, the model can more effectively handle longer code sequences within its token limit~\cite{jana2023attention} ~\cite{dinh2023largelanguagemodelscode}. 

% Once the core translation is completed, the original identifiers can be mapped back into the translated code, ensuring that the final output maintains its functional and structural integrity. By simplifying the task for the LLM, it reduces the number of tokens required to represent the code, thereby reducing computational cost of inference significantly.

\noindent The following are our contributions in this work.

\begin{enumerate}[leftmargin=*]

\item We introduce a novel zero-shot code translation technique that substitutes lengthy, complex user-defined identifiers with concise, generalized placeholders. This method substantially decreases token count and memory consumption, leading to a more efficient and cost-effective process for translating long code with LLMs.

\item The empirical evidence demonstrates that our proposed solution effectively handles long code sequences while maintaining both syntactical and hierarchical information during translation.

\end{enumerate}

% \section{Related Work}
% \input{Sections/rel}

% \subsection{Compilation}
% Data has been taken from the CodeNet~\cite{puri2021codenetlargescaleaicode} and XCodeEval~\cite{khan2024xcodeeval} datasets. Data was partitioned into buckets of code having >2k, >4k, and >8k tokens.
% \vspace{-1.5cm}
\section{Related Work}
Since our work mainly investigates deep metric learning based long code translation approaches, we now first discuss the recent development in machine translation literature, and
then follow it up with how these models are adapted particularly for code translation, a task
which we, in fact, address in this paper.

\para{Generic Machine Translation}
Machine translation (MT) has been an active research area for decades, evolving from rule-based approaches to modern deep learning-based systems. Traditional statistical MT models~\cite{och2003minimum} used phrase-based techniques, but they struggled with maintaining long-range dependencies. The advent of Neural Machine Translation (NMT)~\cite{bahdanau2014neural} introduced attention mechanisms, significantly improving translation quality. Later, the Transformer model~\cite{NIPS2017_3f5ee243} revolutionized MT by efficiently handling long sequences through self-attention. Recent works have extended NMT for various domains, including document-level translation \citep{maruf2021survey}, where maintaining global coherence is critical. However, these models often suffer from ``context truncation issues'' when handling lengthy documents, a limitation similar to what we observe in long code translation. Addressing this, recent research has explored context-aware translation~\citep{voita-etal-2018-context}, yet these approaches remain underexplored in code-related tasks.

\para{Code Translation}
Automated code translation is essential for software portability and legacy system modernization. Early approaches relied on rule-based translation, which required extensive manual effort. With the rise of machine learning, statistical methods \citep{koehn2003statistical} and deep learning-based techniques~\citep{post2024evaluation} have been employed to improve translation accuracy. Recent works have leveraged the pretrained knowledge of language for code translation, considering it as a sequence-to-sequence task. Models such as TransCoder~\cite{roziere2020unsupervised}, CodeT5~\cite{wang2021codet5}, and CodeBERT~\citep{feng-etal-2020-codebert} have demonstrated strong performance in multilingual code translation. Furthermore, execution-based evaluation has been proposed as an alternative to BLEU scores, ensuring functional correctness~\citep{kulal2019spoc}. Despite these advancements, LLMs face severe context-length limitations when dealing with large codebases. This has prompted research into chunking strategies \citep{dikert2016challenges} and hierarchical decoding \citep{zhou2022summarizing} to improve translation efficiency. However, these techniques often introduce fragmentation, causing loss of cross-file dependencies in large code repositories.

\para{Long Code Translation}
Long code translation presents unique challenges, particularly for LLMs constrained by finite context windows. Recent research has explored hierarchical attention mechanisms~\citep{zhou2022summarizing}, external memory augmentation~\citep{wang2024augmenting}. Although these methods improve efficiency, they often require specialized fine-tuning or additional memory resources, making them computationally expensive. Some works focus on \textit{segmenting code intelligently} using \textit{AST-based chunking}~\citep{lin2021improving}, but this still does not resolve the problem of `identifier fragmentation', where long identifiers are split across multiple chunks, degrading model performance. Alternative methods like `syntax-guided translation' \citep{liurepobench} attempt to retain structural consistency, yet those frameworks fail to optimize memory usage effectively. A promising direction in this space is token reduction techniques. For instance, \citet{pan2023tokenize} proposed a prompt compression strategy for reducing token count in LLM-based translation. However, such methods primarily focus on natural language tasks and are not optimized for structured code translation.

Unlike prior chunking-based LCT methods that introduce fragmentation, we propose a zero-shot identifier replacement strategy that directly reduces token count without altering code semantics. While existing models require fine-tuning or additional memory resources, our method is model-agnostic and can be applied to any pretrained LLM. By dynamically substituting long identifiers with compact placeholders, our approach significantly reduces computational cost, making it ideal for industry-scale code translation tasks. To the best of our knowledge, this is the first study to introduce an identifier-aware preprocessing framework for cost-efficient long code translation.

\section{Methodology}
\label{sec:method}
\begin{algorithm}[!h]
\scriptsize
\caption{Process Code and Replace Identifiers}
\label{algo:identifier_replacement}
\KwIn{Code file name, programming language, optional dataset name, optional target directory}
\KwOut{Mapping of original identifiers to new identifiers, modified code}

\SetKwFunction{FMain}{Identifier\_Extraction}
\SetKwFunction{FIdSet}{GetIdSet}
\SetKwFunction{FProcess}{Process\_Code}

\SetKwProg{Fn}{Function}{:}{}

\Fn{\FMain{code, language}}{
    LANGUAGE $\leftarrow$ \textsc{GetLanguage}($language$) \Comment{\hlt{Retrieve programming language metadata}}\\
    parser $\leftarrow$ \textsc{GetParser}($language$)\; \Comment{\hlt{Initialize a syntax parser}}\\
    tree $\leftarrow$ parser.parse(\textsc{Encode}($code$)) \Comment{\hlt{Source code -> Syntax tree}}\\
    root\_node $\leftarrow$ tree.root\_node\; \Comment{\hlt{Extract root node from syntax tree}}\\

    grammar $\leftarrow$ \textsc{LoadGrammar}(language)\; \Comment{\hlt{Load predefined grammar}}\\
    acceptable\_parents $\leftarrow$ grammar[\texttt{`PARENT'}]\; \Comment{\hlt{Retrieve valid parent nodes for identifiers}}\\
    unacceptable\_parents $\leftarrow$ grammar[\texttt{`NOT'}]\; \Comment{\hlt{Retrieve invalid parent nodes for filtering}}\\

    query $\leftarrow$ LANGUAGE.query(\texttt{``(identifier) @identifier"})\; \Comment{\hlt{Query to extract identifiers}}\\
    qc $\leftarrow$ query.captures(root\_node)\; \Comment{\hlt{Get matching identifiers}}\\
    captures $\leftarrow$ [ ]\; \Comment{\hlt{Initialize an empty list for extracted identifiers}}\\

    \ForEach{capture in qc}{
        node $\leftarrow$ capture[0]\;\Comment{\hlt{Extract identifier node}}\\
        parent $\leftarrow$ node.parent\; \Comment{\hlt{Extract the parent node of the identifier}}\\

        \If{parent.type \textbf{not in} acceptable\_parents \textbf{or} code[node.start\_byte:node.end\_byte] \textbf{in} unacceptable\_parents}{
            \textbf{continue}\; \Comment{\hlt{Skip if parent invalid or identifier predefined}}
        }
        \textsc{Append}(captures, capture)\; \Comment{\hlt{Store valid identifiers}}
    }
    \Return captures\; \Comment{\hlt{Return the list of extracted identifiers}}
}

\Fn{\FIdSet{code, captures}}{
    id\_set $\leftarrow$ \textsc{CreateSet}()\; \Comment{\hlt{Initialize an empty set for unique identifiers}}
    \ForEach{capture in captures}{
        node $\leftarrow$ capture[0]\; \Comment{\hlt{Extract identifier node}}\\
        \textsc{Add}(id\_set, code[node.start\_byte:node.end\_byte]) \Comment{\hlt{Add to set}}
    }
    \Return id\_set\; \Comment{\hlt{Return the set of unique identifiers}}
}

\Fn{\FProcess{file\_name, language, dataset, target\_dir}}{
    code $\leftarrow$ \textsc{LoadData}(file\_name, dataset)\; \Comment{\hlt{Load source code from file}}\\
    language $\leftarrow$ \textsc{GetLang}(language)\; \Comment{\hlt{Retrieve language metadata}}
    
    id\_list $\leftarrow$ \textsc{Identifier\_Extraction}(code, language)\; \Comment{\hlt{Extract identifiers from the code}}\\
    identifiers $\leftarrow$ \textsc{GetIdSet}(code, id\_list)\; \Comment{\hlt{Retrieve unique identifiers}}

    i $\leftarrow 0$\; \Comment{\hlt{Initialize index counter}}\\
    mapping $\leftarrow$ \{\}\; \Comment{\hlt{Initialize mapping dictionary}}\\
    len\_reduced $\leftarrow$ 0\; \Comment{\hlt{Initialize length reduction counter}}

    \ForEach{identifier in identifiers}{
        candidate $\leftarrow$ \texttt{id\_}\texttt{i}\; \Comment{\hlt{Generate shorter placeholder for identifier}}

        \If{\textsc{TOKEN\_Length}(candidate) $\geq$ \textsc{TOKEN\_Length}(identifier)}{
            mapping[identifier] $\leftarrow$ identifier\; \Comment{\hlt{Keep original identifier if replacement is not shorter}}
        }
        \Else{
            len\_reduced $\leftarrow$ len\_reduced + \textsc{TOKEN\_Length}(identifier) - \textsc{TOKEN\_Length}(candidate)\;\\
            mapping[identifier] $\leftarrow$ candidate\; \Comment{\hlt{Replace identifier with a shorter placeholder}}\\
            i $\leftarrow$ i + 1\; \Comment{\hlt{Increment placeholder index}}
        }
    }

    \ForEach{(old, new) in mapping}{
        \If{old \textbf{in} [\texttt{``file\_name"}, \texttt{``language"}, \texttt{``len\_reduced"}, \texttt{``dataset"}]}{
            \textbf{continue}\; \Comment{\hlt{Skip protected keywords from replacement}}
        }
        code $\leftarrow$ \textsc{Replace}(code, old, new)\; \Comment{\hlt{Replace original identifier with its mapped value}}
    }
    \Return mapping, code\; \Comment{\hlt{Return updated mapping and modified code}}
}
\end{algorithm}

\begin{figure*}[t]
\centering
\includegraphics[width=.99\textwidth]{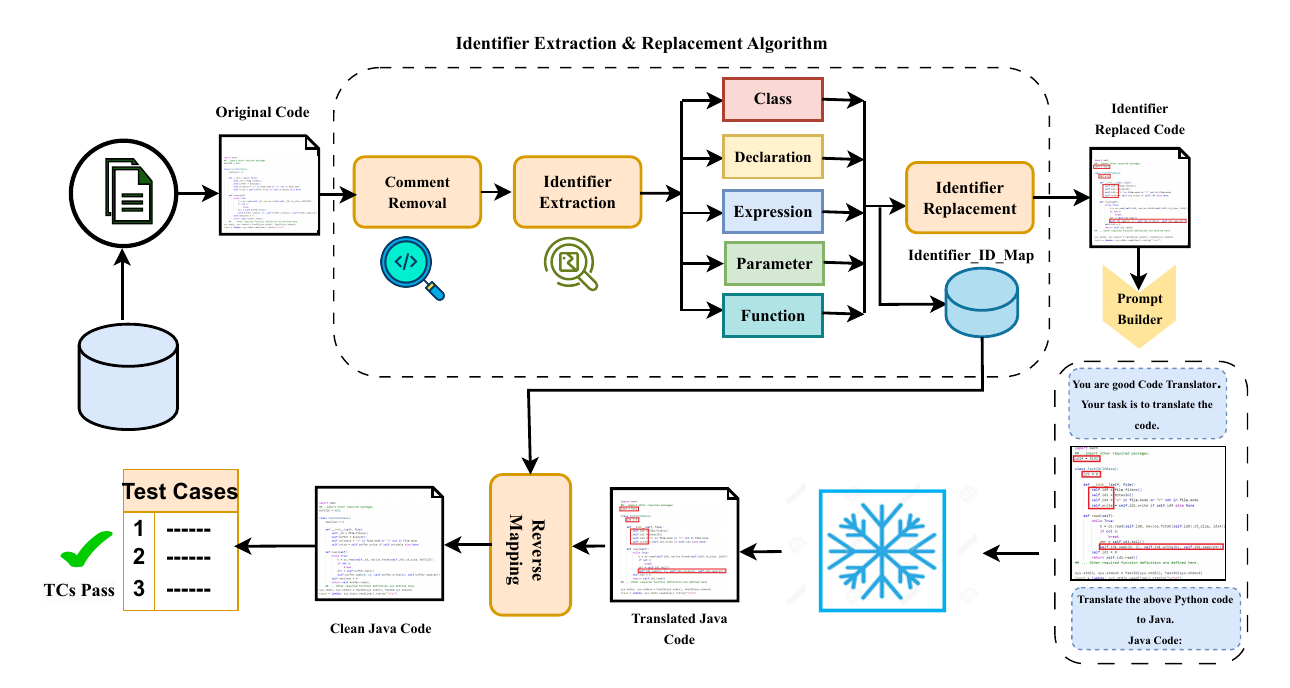}
\caption{An overview of the identifier extraction and replacement algorithm for long source code translation. The process involves identifier extraction, classification into syntactic categories, replacement using an identifier mapping strategy, code translation, and final restoration to ensure syntactic and semantic correctness.}
\label{fig:lct_pic}
\end{figure*}

% In this section we define the task description and delve into a comprehensive explanation of our proposed long code translation framework using identifier replacement.

In this section, we outline the task description and provide a detailed explanation of our proposed approach as depicted in the Figure~\ref{fig:lct_pic}, to perform long code translation with LLMs through the use of identifier replacement.
% \vspace{-2.2cm}
\para{Task Description} The task of translating source code in a zero-shot scenario with a LLM $\mathcal{M}$ can be described as using a multinomial sampling strategy to convert code written in a source language $\mathcal{L}_s$ into corresponding code in a target language $\mathcal{L}_t$.
% Source code translation task in zero-shot scenario can be defined as applying multinomial sampling strategy using a LLM $\mathcal{M}$ to map source code written in a language $\mathcal{L}_s$ to target code in another language $\mathcal{L}_t$. 
Formally speaking, given an input code sequence $C_s = \{x_1, x_2, \ldots, x_n\}$ in $\mathcal{L}_s$, the $\mathcal{M}$ generates the output sequence $C_t = \{y_1, y_2, \ldots, y_m\}$ in $\mathcal{L}_t$, such that the functionality of $C_s$ is preserved.

To mitigate context length limitations, we utilize an \textit{identifier extraction and replacement} strategy during translation.  First, we identify the set of unique long identifiers, $\mathcal{I}_s = \{i_1, i_2, \dots, i_k\}$, within the source code $C_s$. These identifiers are then replaced with a set of generalized placeholders, $\mathcal{P} = \{p_1, p_2, \ldots, p_k\}$, resulting in a simplified and shorter source sequence $C'_s = \{x'_1, x'_2, \dots, x'_n\}$. The LLM translates $C'_s$ in the target language $\mathcal{L}_t$. Finally, in a post-processing step, the placeholders identifiers $\mathcal{P}$ are mapped back to corresponding 
original identifiers $\mathcal{I}_s$ within the translated code $C_t$, preserving the code's functionality and semantics.

\para{Identifier Extraction and Replacement} In this method as described in the Algorithm 1, we employ tree-sitter\cite{tree_sitter}, a powerful parsing tool, to systematically extract relevant long identifiers from source code. By generating precise syntax trees across multiple programming languages, tree-sitter allows us to identify key programming constructs, which serve as primary candidates\footnote{A detailed statistical overview of the probable identifier list is provided in Table\ref{tab:identifier_classes}} for identifier replacement. The following key programming constructs are the main candidates used for identifier extraction and replacement. 

% \subsubsection{Syntax Tree Parsing with Tree-sitter}
% Identifier extraction plays a pivotal role in preparing the long source code for subsequent translation processes.

% \textbf{Selection of Target Node:} 

% Initially, the source code undergoes parsing using tree-sitter, focusing on identifying nodes corresponding to key programming constructs such as \textit{function definitions, variable declarations}, \textit{parameter lists}, and \textit{class definitions}. 

    \begin{itemize}
    \item \textbf{Function Definitions:} To capture function names and associated parameter identifiers.
    \item \textbf{Variable Declarations:} To extract variable names, ensuring the preservation of variable semantics across the translation process.
    \item \textbf{Class and Method Declarations:} To capture class names and method signatures, crucial for maintaining object-oriented structure.
    \item \textbf{Reserved Identifiers: } Identifiers such as \textit{this}, \textit{self}, \textit{super}, \textit{null}, \textit{true}, and \textit{false} are not replaced to maintain language-specific semantics. Additionally, built-in function names such as \textit{print} (Python), \textit{System.out.println} (Java), \textit{std::cout} (C++), and keywords like \textit{def}, \textit{class}, \textit{static}, and \textit{return} are preserved to ensure syntactic correctness.
\end{itemize}
Employing these key programming constructs, we developed grammars for each programming language which are then used to extract the potential identifiers suitable for replacements.
Let $\mathcal{I}_s = \{i_1, i_2, \ldots, i_k\}$ represent the set of extracted identifiers. For each $i_j \in \mathcal{I}_s$, we assign a unique placeholder $p_j \in \mathcal{P}$ of a small token length = $2$ and produce a simplified source code $C'_s$ from original source code $C_s$. This way, we achieve the length reduction, $\Delta l$, in the token length as  given by: 
$\Delta l = \sum_{j=1}^{k} \left( |i_j| - |p_j| \right)$, where $|i_j|$ and $|p_j|$ denote the token length of the original identifier and its corresponding placeholder, respectively.
% This transformation simplifies the input for the LLM by reducing the number of unique tokens it has to process. 

After translation is complete and the LLM generates the output sequence $C'_t$ in the target language, a reverse process is applied to restore the original identifiers. Let $g$ represent the reverse mapping function, which takes translated sequence $C'_t$ and original set of identifiers $\mathcal{I}_t$ in the target language, and produces the final translated sequence $C_t$:\[
C_t = g(C'_t, \mathcal{I}_t) = \{y_1, y_2, \dots, y_m\}\]
\[\quad g : C'_t \times \mathcal{I}_t \rightarrow C_t\]
This post-processing step ensures that placeholders in the translated code are replaced with the corresponding identifiers from the target language, preserving the semantics of the original source code. 

% \para{Zero-Shot Inference}
% In the zero-shot setting, the LLM $\mathcal{M}$ is prompted to generate the target code sequence $C'_t$ from the simplified source code sequence $C'_s$. This process can be treated as computing the most likely sequence of tokens in the target language $\mathcal{L}_t$ given the input sequence in the source language $\mathcal{L}_s$. Formally, LLM computes the posterior probability of the target sequence as:
% \[
% P(C'_t | C'_s) = \prod_{t=1}^{m} P(y'_t | C'_s, y'_1, y'_2, \dots, y'_{t-1}),
% \]
% where $P(y'_t | C'_s, y'_1, y'_2, \dots, y'_{t-1})$ represents the probability of generating the next token $y'_t$ conditioned on the entire input sequence $C'_s$ and the previously generated tokens $\{y'_1, y'_2, \dots, y'_{t-1}\}$ in the target sequence. The objective of the LLM is to find the sequence $C'_t$ that maximizes the posterior probability:
% \[
% C'_t = \arg \max_{C'_t} P(C'_t | C'_s).
% \]

 \vspace{-0.7cm}
\section{Experimental Setup}
% \vspace{-.4cm}

\begin{figure*}[t]
\centering
\includegraphics[width=.99\textwidth]{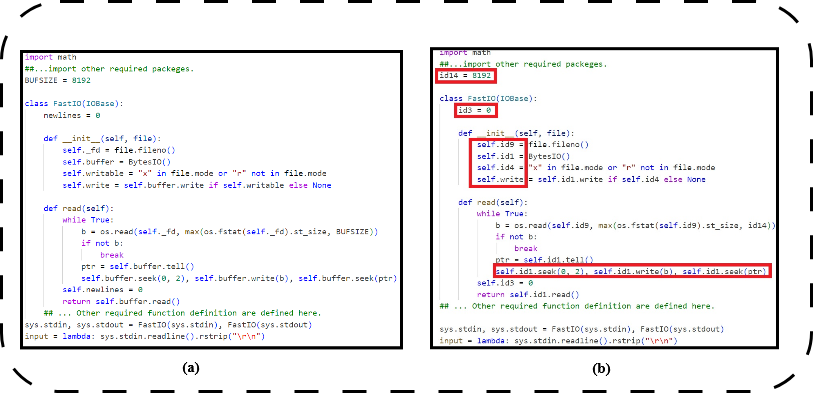}
\caption{The figure illustrates the transformation of source code during the identifier replacement (IdRep) process.  
\textbf{(a)} represents the original source code with long, descriptive identifiers, while  
\textbf{(b)} shows the modified code where identifiers are replaced with shorter placeholders.  
This transformation helps reduce token count, enabling LLMs to process longer sequences efficiently while preserving semantic correctness.}
\label{fig:lct_example}
\end{figure*}

In this section, we provide an overview of the dataset used in our experiments. We then explain various closed and open source LLMs, along with their specific configurations, to assess their capacity in translating long code sequences using identifier replacements.

\subsection{Dataset}
\begin{table}
\centering
\small
\begin{adjustbox}{width=.99\columnwidth}
\begin{tabular}{@{}l|ccccc|ccccc@{}}
\toprule
\multicolumn{8}{c}{\textbf{Original Dataset}} & \multicolumn{2}{c}{\textbf{Reduced Dataset}} \\ \midrule
\textbf{Length} & \textbf{C} & \textbf{C++} & \textbf{Java} & \textbf{Go} & \textbf{Python} & \textbf{C} & \textbf{C++} & \textbf{Java} & \textbf{Go} & \textbf{Python} \\ \midrule
2000 & 714 & 90,015 & 568 & 51,418 & 6,596 & 20  & 1,717 & 577 & 33 & 37 \\
4000 & 214 & 19,545 & 70  & 11,812 & 2,221 & 7   & 416   & 109 & 4  & 17 \\
8000 & 22  & 4,134  & 3   & 2,048  & 258 & 1   & 162   & 22  & 0  & 4  \\
\bottomrule
\end{tabular}
\end{adjustbox}
\caption{Statistics of XcodeEval dataset. The table shows the total number of code instances across five programming languages (C, C++, Java, Go, Python) in both the Original and Reduced subsets.}
\label{tab:data_comparison}
\end{table}

% We evaluated our approach using the XcodeEval dataset \cite{khan2024xcodeeval}, a subset of CodeNet \cite{puri2021codenetlargescaleaicode}. XcodeEval comprises C, C++, Java, Go, and Python code instances with associated test cases, enabling the evaluation of cross-language code translation accuracy. 

% We use the XcodeEval dataset~\cite{khan2024xcodeeval}, derived from CodeNet~\cite{puri2021codenetlargescaleaicode} for the experiments. The dataset consists of instance for five programming languages: C, C++, Java, Go, and Python and without corresponding test cases to evaluate the correctness of the translation. 
% Our experiments utilize the XcodeEval dataset~\cite{khan2024xcodeeval}, derived from CodeNet~\cite{puri2021codenetlargescaleaicode} and XcodeEval datasets, covering five programming languages: C, C++, Java, Go, and Python. 

To analyze our approach of code translation via identifier replacement, we partitioned the dataset into three subsets based on token length: those with lengths greater than or equal to 2000, 4000, and 8000 tokens. These lengths were determined using the TikToken tokenizer~\cite{pan2023tokenize}. We also excluded instances without corresponding test cases to ensure accurate assessment. Table\ref{tab:data_comparison} presents the total number of source code instances across various languages for both the original and reduced subsets. The "Original Dataset" columns show raw counts before filtering, while the "Reduced Dataset" columns indicate the instances retained after filtering.

\subsection{LLM Models}
In this work, we employ models such as GPT-3.5-Turbo \cite{ye2023comprehensive}, GPT4o-Mini \cite{hurst2024gpt}, Mixtral (8×7B) \cite{jiang2023mistral}, and CodeLlama \cite{grattafiori2023code}. The choice of these models is primarily driven by their widespread adoption, popularity, and the constraints of available computational resources. Although some of these models can theoretically handle inputs exceeding the token limits of the dataset mentioned earlier, we can still assess the effectiveness of our identifier replacement technique for long code translation tasks.

\section{Results \& Analysis}

% In this section we present the experimental results and key observations organized within pivotal research questions (\textbf{RQ}) for long code translations using identifier replacements. 
% 
\begin{table*}
\centering
\small
\begin{adjustbox}{width=.99\textwidth}
\begin{tabular}{@{}l@{~~~}l@{~~~}c| ccc| ccc| ccc| ccc@{}}
\toprule
\multirow{2}{*}{} & \multirow{2}{*}{\textbf{Language}} & \multirow{2}{*}{\textbf{\# of Samples}} &
\multicolumn{3}{c}{\textbf{GPT3.5-Turbo}}             & \multicolumn{3}{c}{\textbf{GPT 4o-Mini}} & \multicolumn{3}{c}{\textbf{Mixtral 8*7B}} & \multicolumn{3}{c}{\textbf{CodeLlama}} \\
\cmidrule(r){4-6} \cmidrule(r){7-9} \cmidrule(r){10-12} \cmidrule(r){13-15}
& & & \textbf{NoRep}      & \textbf{IdRep}      &  & \textbf{NoRep}     & \textbf{IdRep}     &  & \textbf{NoRep} & \textbf{IdRep} &  & \textbf{NoRep} & \textbf{IdRep} &  \\
\midrule

\multirow{5}{*}{\rotatebox{90}{2000}} 
& C & 20 & 37.1\% & \textbf{37.5\%} &  & 41.0\% & \textbf{45.7\%} &  & 10.3\% & \textbf{11.0\%} &  & 12.1\% & \textbf{13.0\%} & \\
& C++ & 1717 & \textbf{10.5}\% & 10.4\% &  & \textbf{17.0\%} & 14.5\% &  & 2.6\% & \textbf{2.6\%} &  & \textbf{3.6\%} & 3.3\% & \\
& Java & 577 & 14.3\% & \textbf{15.0\%} &  & \textbf{21.3\%} & 20.5\% &  & \textbf{4.3\%} & 3.6\% &  & 3.9\% & \textbf{4.1\%} & \\
& Go & 33 & 35.5\% & \textbf{38.0\%} &  & 58.5\% & \textbf{61.3\%} &  & 7.6\% & \textbf{8.0\%} &  & 8.3\% & \textbf{8.8\%} & \\
& Python & 37 & 32.8\% & \textbf{35.0\%} &  & 33.3\% & \textbf{40.0\%} &  & 8.6\% & \textbf{9.0\%} &  & 9.3\% & \textbf{10.3\%} & \\
\midrule

\multirow{5}{*}{\rotatebox{90}{4000}} 
& C & 7 & 22.5\% & \textbf{24.5\%} &  & 41.0\% & \textbf{41.3\%} &  & \textbf{5.8\%} & 5.4\% &  & - & - & \\
& C++ & 416 & 14.0\% & \textbf{16.0\%} &  & \textbf{25.0\%} & 20.3\% &  & \textbf{2.8\%} & 1.4\% &  & - & - & \\
& Java & 109 & \textbf{11.8\%} & 10.9\% &  & \textbf{17.5\%} & 14.5\% &  & \textbf{2.3\%} & 2.0\% &  & - & - & \\
& Go & 4 & - & - &  & - & - &  & - & - &  & - & - & \\
& Python & 17 & 20.3\% & \textbf{28.6\%} &  & 32.8\% & \textbf{35.0\%} &  & 5.3\% & \textbf{5.8\%} &  & - & - & \\
\midrule

\multirow{5}{*}{\rotatebox{90}{8000}} 
& C & 1 & - & - &  & - & - &  & - & - &  & - & - & \\
& C++ & 162 & 7.5\% & \textbf{9.8\%} &  & 15.8\% & \textbf{16.5\%} &  & - & - &  & - & - & \\
& Java & 22 & \textbf{7.4\%} & 6.8\% &  & 16.0\% & \textbf{17.8\%} &  & - & - &  & - & - & \\
& Go & 0 & - & - &  & - & - &  & - & - &  & - & - & \\
& Python & 4 & - & - &  & - & - &  & - & - &  & - & - & \\
\bottomrule
\end{tabular}
\end{adjustbox}
\caption{Performance comparison on the XcodeEval dataset across the LLMs, under two configurations: No Replacement (NoRep) and Identifier Replacement (IdRep). The table is organized by sample size (2000, 4000, and 8000) and five programming languages (C, C++, Java, Go, and Python). Performance is evaluated using accuracy metric. The final performance is computed over the average for each source language. The best-performing results for each setting are highlighted in bold. Due to insufficient samples (less than 5 instances), we do not report the average accuracy for certain languages. Additionally, due to context length limitations, we do not report results for Mixtral and CodeLlama for the 4000 and 8000 token length buckets.}
\label{tab:results}
\end{table*}
In this section, we present the experimental results and significant observations organized around key research questions (RQ) related to long code translations using identifier replacements. To evaluate our identifier replacement approach (\textbf{IdRep}) for long code translation, we establish a baseline using translations without identifier replacement (\textbf{NoRep}). Additionally, an example comparing the IdRep and NoRep methods is depicted in Figure~\ref{fig:lct_example}. A translation is deemed successful only if it compiles, executes, \emph{and} passes all unit tests provided by a particular dataset. Our evaluation indeed covers both syntactic correctness and functional behavior. We therefore emphasize that our evaluation validates functionality under the available test suite, but broader runtime behavior remains an open challenge for future work. Thus, in this work we investigate the following key research questions.

% , aligning with the objectives outlined in this section. 

% As discussed in Section~\ref{sec:intro} that the LCT is a non-trivial task as compared to traditional code translation task. Thus, in this work we investigate the following research questions.

\uls
\li \textbf{RQ-1}: What is the impact of identifier replacements on translation performance for long code across different programming languages, LLM models, and different token length buckets?
\ule

\uls
\li \textbf{RQ-2}: To what extent do identifier replacements reduce token length in LLM-based code translation across different programming languages?
\ule

\begin{table}
\centering
\renewcommand{\arraystretch}{1} % 
\adjustbox{max width=.8\columnwidth}{
\begin{tabular}{@{}ccc@{}}
\toprule
\textbf{Language} & \textbf{Token Saved (per sample)}\\
% & \textbf{Token Saved (total)} \\ 
\midrule
C       & 22.76\\ 
% & 19,248\\ 
Java    & 11.34\\
% & 583,197\\
Go      & 17.21\\
% & 9,777\\
C++     & 5.75\\
% & 517,898\\
Python  & 15.29\\
% & 100,875\\ 
\midrule
\end{tabular}
}
\caption{This table presents average number of tokens saved per sample across different programming languages in the XcodeEval dataset when applying the IdRep strategy. The reduction in token count enables more efficient utilization of LLM context windows, benefiting languages with extensive identifier usage.}
\label{tab:merged_token_cost}
\end{table}

\para{Identifier replacement vs. language} The translation accuracies with and without identifier replacements (\textit{IdRep} and \textit{NoRep}, respectively) for various programming languages, token length buckets and LLMs are presented in the Table~\ref{tab:results}. We observed that the \textbf{Identifier replacement, as a context length reduction strategy during translation, tends to be more effective for procedural programming languages (e.g., C, Go) than object-oriented languages (e.g., C++, Java)}. For instance, Identifier replacement in the 2000-token bucket yields contrasting results for GPT-4o model where translation accuracy improves significantly for C (41.0\% to 45.7\%), while the performance decreases for C++ (17.0\% to 14.5\%). This can be due to relatively simpler structure of procedural programming languages compared to object-oriented languages, where  meaning of an identifier is highly context-dependent. By substituting these identifiers with abstract placeholders, the model risks losing vital contextual information necessary for maintaining code correctness, which could potentially disrupt object-oriented dependencies.

Table~\ref{tab:merged_token_cost} illustrates the efficacy of identifier replacement in reducing token counts across various programming languages, enabling larger code segments to be accommodated within the LLM context window for translation. Thus, replacing identifiers can significantly reduce translation costs for long code sequences when using LLMs. This is particularly beneficial for translating complex industry-specific codebases that often use lengthy, user-defined identifiers.

\para{Identifier replacement vs. LLMs} In the Table ~\ref{tab:results}, we observed an inconsistency in improvements across models which indicates that
\textbf{identifier replacement benefits the code translation performance of larger models (e.g., GPT) more than smaller models (e.g., Mixtral, CodeLlama)}. This suggests weaker abstraction in smaller models and a greater reliance on explicit semantic cues for accurate long code translation.
Although the average reduction may appear modest in relative terms, it is important to note that this reduction often targets the most critical bottleneck positions long user-defined identifiers. These identifiers not only inflate token counts but also frequently exceed subword segmentation boundaries, leading to poor generalization by LLMs. By systematically replacing such high-cost tokens, our method creates space for including additional lines of logic that would otherwise be truncated. Thus, the benefit is not only in percentage reduction but in \textbf{strategically freeing context window capacity where it matters most.}

\section{Ablation Study}

% To further investigate the impact of identifier semantics on long code translation, we conducted an ablation study where we selectively replaced identifiers belonging to specific syntactic categories. The objective of this study is to analyze how identifier replacement affects translation performance when applied to different structural components of a program, such as function names, variable declarations, class names, etc.

We conducted an ablation study to isolate the impact of identifier semantics on long code translation.  We selectively replaced identifiers in different syntactic categories to analyze how this affects translation performance based on the identifiers' structural role.
% Through an extensive dataset analysis, 
We categorized identifiers into five major categories (as shown in Table~\ref{tab:identifier_classes}): \textit{function definitions, variable declarations, class-related elements, method names}, and \textit{function parameters}. These categories represent key elements in program structure and logic. This study used the GPT-4o-Mini model with a 2000-token length bucket to translate Java and C++ codes to other languages.

% \footnote{It is to be noted that when we applied a particular identifier replacement strategy to other token-length buckets (4000, 8000), we did not obtain a sufficient number of dataset instances to conduct additional analyses. Additionally, in the 2000-token bucket, for individual class identifier replacements, there were not enough instances for Java and C++. Similarly, in the 2000-token bucket, when performing individual identifier replacements, we did not obtain a significant number of instances for Python, Go, and C languages to conduct further analysis.}
\begin{table}[t]
    \centering
    \small
    \begin{adjustbox}{width=.95\columnwidth}
    \begin{tabular}{|c|l|}
        \hline
        \textbf{Category} & \textbf{Code Elements} \\
        \hline
        \multirow{6}{*}{Function (F)} 
        & \texttt{compact\_constructor\_declaration} \\
        & \texttt{constructor\_declaration} \\
        & \texttt{method\_declaration} \\
        & \texttt{function\_declaration} \\
        & \texttt{function\_definition} \\
        & \texttt{function\_declarator} \\
        \hline
        \multirow{6}{*}{Declaration (D)} 
        & \texttt{variable\_declarator} \\
        & \texttt{declaration} \\
        & \texttt{array\_declarator} \\
        & \texttt{var\_spec} \\
        & \texttt{initializer\_pair} \\
        & \texttt{pointer\_declarator} \\
        \hline
        \multirow{5}{*}{Class (C)} 
        & \texttt{class\_definition} \\
        & \texttt{class\_declaration} \\
        & \texttt{interface\_declaration} \\
        & \texttt{enum\_declaration} \\
        & \texttt{annotation\_type\_declaration} \\
        \hline
        \multirow{5}{*}{Expression (E)} 
        & \texttt{assignment\_expression} \\
        & \texttt{initializer\_list} \\
        & \texttt{assignment} \\
        & \texttt{parenthesized\_expression} \\
        & \texttt{array\_initializer} \\
        \hline
        \multirow{7}{*}{Parameter (P)} 
        & \texttt{optional\_parameter\_declaration} \\
        & \texttt{typed\_parameter} \\
        & \texttt{parameter\_declaration} \\
        & \texttt{receiver\_parameter} \\
        & \texttt{formal\_parameter} \\
        & \texttt{lambda\_parameters} \\
        & \texttt{default\_parameter} \\
        \hline
    \end{tabular}
    \end{adjustbox}
     \caption{The categorization of Identifiers in the source Code. This table classifies identifiers into five major categories based on their syntactic roles in source code: Function (F) for function and method definitions, Declaration (D) for variable and construct declarations, Class (C) for class and interface declarations, Expression (E) for assignment and initialization expressions, and Parameter (P) for function and method parameters.}
     \label{tab:identifier_classes}
\end{table}

\begin{table}
\centering

\begin{adjustbox}{width=.95\columnwidth}
\begin{tabular}{@{}l@{~~~}l@{~~~}ccc|ccc@{}}
\toprule
\multirow{2}{*}{} & \multirow{2}{*}{} &
\multicolumn{3}{c|}{\textbf{Java}} & \multicolumn{3}{c}{\textbf{C++}} \\
\cmidrule(r){3-5} \cmidrule(r){6-8}
& & \textbf{Acc.} & \textbf{NoRep Acc.} & \textbf{IdRep Acc.} 
& \textbf{Acc.} & \textbf{NoRep Acc.} & \textbf{IdRep Acc.} \\
\midrule

\multirow{4}{*}{}
& Declaration & 19.7\% & \multirow{4}{*}{21.3\%} & \multirow{4}{*}{20.5\%} & 16.7\% & \multirow{4}{*}{17.0\%} & \multirow{4}{*}{14.5\%} \\
& Function & 22.2\% & & & 16.7\% & & \\
& Parameter & 16.0\% & & & {\bf 23.5}\% & & \\
& Expression & {\bf 22.7}\% & & & 17.7\% & & \\
\bottomrule
\end{tabular}
\end{adjustbox}
\caption{Performance comparison of Java and C++ translations across different identifier replacement categories. The table evaluates the translation performance when selectively replacing different identifier types (Declaration, Function, Parameter, and Expression) for Java and C++ while translating them into C, C++, Go, and Python. The accuracy (Acc.) represents the mean accuracy across the target languages for each identifier category, while the IdRep Accuracy (IdRep Acc.) represents the average accuracy when all identifier types are replaced and the NoRep Accuracy (NoRep Acc.) represents the average accuracy without any replacement as reported in Table~\ref{tab:results}. The best performance is in bold.}
\label{tab:comparison}
\end{table}

\textbf{Effective long code translation through identifier replacement necessitates syntactic awareness specific to each programming language}. For instance, when translating from Java, identifier replacement (IdRep) accuracy is 20.5\% when all identifiers are replaced.  However, restricting replacement to only \textit{function} or \textit{expression} identifiers significantly improves accuracy to approximately 22\%. Similarly, translating from C++, selectively replacing only \textit{parameter} identifiers yields the best performance (23.5\%). This significantly outperforms both benchmark translations with no identifier replacement (17\%) and translations with all the identifiers replaced (14.5\%).

Further, our findings show that replacing \textit{declaration} identifiers harms translation performance across programming languages, likely due to their fundamental role in code structure. In contrast, replacing \textit{parameter} and \textit{expression} identifiers significantly improves long code translation when using identifier replacements.

\section{Conclusion and Future Directions}
% In this paper, we introduced a novel zero-shot approach for long code translation using identifier replacement, which addresses key challenges such as context window and memory limitations inherent in LLMs. By replacing longer user-defined identifiers with shorter generalized placeholders, our method reduces the token count and enhances the efficiency of long code translation without compromising the syntactical and functional integrity of the code. The empirical results demonstrate that this approach not only leads to significant cost savings but also improves translation accuracy for few languages and for rest it produces comparable performance. As a future work one hybrid techniques can be proposed that integrate identifier replacement with symbolic execution or program synthesis to further improve performance.

We present a novel zero-shot approach for long code translation that leverages identifier replacement to overcome the context window and memory limitations of LLMs. By substituting long identifiers with shorter placeholders, the method reduces token count, leading to cost-effective and efficient translation without sacrificing code integrity. Our experiments demonstrate significant cost savings and improved translation accuracy. 
Future research can focus on hybrid techniques, such as combining identifier replacement with symbolic execution, to further boost translation performance.

% This paper presents a novel zero-shot approach for long code translation using identifier replacement, addressing LLM limitations like context window and memory constraints. By substituting lengthy identifiers with shorter placeholders, the method reduces token count, boosting translation efficiency without sacrificing syntactical or functional integrity. Empirical results demonstrate significant cost savings and improved translation accuracy in certain languages, with comparable performance in others. Future work could explore hybrid techniques, combining identifier replacement with symbolic execution or program synthesis, to further enhance translation performance.

\section{Limitations}
While our identifier replacement strategy significantly reduces token length and enhances computational efficiency in long code translation, several limitations remain. First, the approach assumes that replacing long identifiers with placeholders does not affect semantic interpretation, but this assumption may not always hold, particularly for programming languages with strong type systems and deep interdependencies between identifiers. In object-oriented languages like Java and C++, identifier replacement may lead to loss of context, affecting method resolution and dependency tracking. Additionally, the effectiveness of identifier replacement varies across programming paradigms, with procedural languages (e.g., C, Go) benefiting more than those with complex inheritance and polymorphism. Another limitation arises in the availability datasets for long codebases to empirically validate our proposed approach. Finally, our evaluation primarily focuses on syntactic correctness, leaving functional correctness and runtime behavior as open challenges that require further investigation.

%\bibliographystyle{ACM-Reference-Format}
\bibliographystyle{plainnat}
\bibliography{sample-base}

\end{document}